\documentclass[reprint,NumberedRefs]{JASA}
\usepackage{enumitem} 
\usepackage{graphicx} 
\usepackage{tabularx}

\begin{document}

\def\a{{\mathbf a}}
\def\d{{\mathbf d}}
\def\e{{\mathbf e}}
\def\n{{\mathbf n}}
\def\p{{\mathbf p}}
\def\r{{\mathbf r}}
\def\s{{\mathbf s}}
\def\v{{\mathbf v}}
\def\w{{\mathbf w}}
\def\x{{\mathbf x}}
\def\y{{\mathbf y}}
\def\z{{\mathbf z}}

\def\0{{\mathbf 0}}
\def\1{{\mathbf 1}}

\def\A{{\mathbf A}}
\def\iC{{\mathbf C}}
\def\B{{\mathbf B}}
\def\D{{\mathbf D}}
\def\E{{\mathbf E}}
\def\G{{\mathbf G}}
\def\H{{\mathbf H}}
\def\I{{\mathbf I}}
\def\J{{\mathbf J}}
\def\K{{\mathbf K}}
\def\M{{\mathbf M}}
\def\P{{\mathbf P}}
\def\Q{{\mathbf Q}}
\def\R{{\mathbf R}}
\def\S{{\mathbf S}}
\def\X{{\mathbf X}}
\def\Y{{\mathbf Y}}
\def\P{{\mathbf P}}
\def\W{{\mathbf W}}
\def\Z{{\mathbf Z}}

\def\bmu{{\boldsymbol{\mu}}}
\def\btheta{{\boldsymbol{\theta}}}
\def\bgamma{{\boldsymbol{\gamma}}}
\def\blambda{{\boldsymbol{\lambda}}}
\def\btheta{{\boldsymbol{\theta}}}
\def\balpha{{\boldsymbol{\alpha}}}
\def\bxi{{\boldsymbol{\xi}}}
\def\bzeta{{\boldsymbol{\zeta}}}

\def\bGamma{{\boldsymbol{\Gamma}}}
\def\bLambda{{\boldsymbol{\Lambda}}}
\def\bSigma{{\boldsymbol{\Sigma}}}

\def\cC{{\cal C}}
\def\F{{\cal F}}
\def\N{{\cal N}}
\def\L{{\cal L}}
\def\U{{\cal U}}

\def\cT{{\mathcal T}}
\def\cA{{\mathcal A}}

\def\C{{\mathbb C}}
\def\bM{{\mathbb M}}
\def\bD{{\mathbb D}}
\def\bE{{\mathbb E}}
\def\bR{{\mathbb R}}
\def\bS{{\mathbb S}}
\def\bU{{\mathbb U}}
\def\bV{{\mathbb V}}

\title[JASA/Sample JASA Article]{Localization of DOA trajectories~--~Beyond the grid}

\author{Ruchi Pandey}
\email{ruchi.pandey@research.iiit.ac.in}
\author{Santosh Nannuru}
\email{santosh.nannuru@iiit.ac.in}
\affiliation{Signal Processing and Communications Research Center, IIIT Hyderabad, India}


\date{\today} 
\begin{abstract}
The direction of arrival (DOA) estimation algorithms are crucial in localizing acoustic sources. Traditional localization methods rely on block-level processing to extract the directional information from multiple measurements processed together. However, these methods assume that DOA remains constant throughout the block, which may not be true in practical scenarios. Also, the performance of localization methods is limited when the true parameters do not lie on the parameter search grid. 
In this paper we propose two trajectory models, namely the polynomial and bandlimited trajectory models, to capture the DOA dynamics. To estimate trajectory parameters, we adopt two gridless algorithms: i) Sliding Frank-Wolfe (SFW), which solves the Beurling LASSO problem and ii) Newtonized Orthogonal Matching Pursuit (NOMP), which improves over OMP using cyclic refinement. Furthermore, we extend our analysis to include wideband processing. The simulation results indicate that the proposed trajectory localization algorithms exhibit improved performance compared to grid-based methods in terms of resolution, robustness to noise, and computational efficiency.
\end{abstract}
\maketitle
\section{\label{sec:1} Introduction}
The recent advancements in robotics and autonomous devices have led to a growing demand for reliable and efficient localization and tracking algorithms \cite{alenljung2019user,han2015two,okutani2012outdoor,nakamura2011intelligent}. Applications such as smart devices and hearing aids require algorithms that can accurately determine the direction of arrival (DOA) of acoustic sources in real-time \cite{wan2016application,farmani2017informed}. The field of localization algorithms is abundant with a range of techniques. Conventional beamforming (CBF) is computationally efficient but lacks resolution when multiple sources are close to each other \cite{van1988beamforming}. On the other hand, subspace-based methods, such as multiple signal classification (MUSIC) and its variants, offer improved resolution but require more data snapshots \cite{music1986,rao1989performance}. There are also sparse recovery methods like least absolute shrinkage and selection operator (LASSO) \cite{chen2001atomic,compressivebeamforming,l1svd}, orthogonal matching pursuit (OMP) \cite{omp,pati1993orthogonal}, and sparse Bayesian learning (SBL) \cite{tipping2001,wipf2007}. LASSO is vulnerable to the choice of the regularization parameter, and computational complexity increases with data size. OMP is a greedy algorithm, and multiple coherent sources may impact its performance. 
SBL has the unique property of automatic sparsity selection and does not require regularization \cite{wipf2007,gerstoft2015multiple}. It is a high-resolution method but suffers from basis mismatch and can be computationally expensive, especially when dealing with large dictionaries or high-dimensional data \cite{multifreq2019sparse,gemba2019robust,MLEefficient,pandey2021sparse,pandey2022experimental}.


While the literature contains numerous DOA estimation algorithms, most of them process multiple snapshots together to estimate fixed DOA within a block \cite{krim1996two,van1988beamforming,music1986,vantrees2002,mvdr2005,dibiase2000high,diaz2020robust,pandey2021sparse}. However, in real-world scenarios, the DOA is not constant across the snapshots, which can lead to limitations in the performance of localization algorithms. 
In \cite{park2021sequential}, a sequential SBL algorithm was proposed to estimate time-varying DOAs, while in \cite{opochinsky2021deep,diaz2020robust}, neural network-based methods were used to obtain trajectories directly. Despite these advancements, there is still a need for algorithms that can accurately estimate DOA trajectories while being computationally efficient.

In our previous work \cite{pandey2022parametric,shreyas_TLDL}, we introduced a parametric trajectory model incorporating linear source motion within a block and performed trajectory localization (TL). We developed TL-CBF and TL-SBL algorithms \cite{pandey2022parametric} and deep U-Net architecture \cite{shreyas_TLDL} for estimating the linear parameters, which perform better than traditional methods for moving sources. While TL-SBL has a higher resolution than TL-CBF, the computational cost is far too high for real-time applications, and the performance is compromised when the source trajectories don't lie on the predefined grid. The use of finer grids significantly increases the computational cost.

To address the limitations of grid-based localization algorithms, various gridless methods have been proposed. Gridless localization has been formulated as an atomic norm minimization (ANM) problem and solved using semi-definite programming in 1D and 2D scenarios \cite{tang2013compressed,xu2014precise,xenaki2015grid,bhaskar2013atomic,chi2014compressive,yang2017two,yang2018resolution,zhang2022efficient,yang2016vandermonde,wu2022maximum}. Additionally, gridless methods have been applied for non-uniform arrays and wideband processing \cite{semper2018grid,wagner2021gridless,wu2022gridless,jiang2020gridless,ang2020multiband,chardon2021gridless}. The Newtonized OMP (NOMP) algorithm is a variation of OMP that employs Newton steps to refine source parameters in each iteration \cite{mamandipoor2016newtonized}. 
An alternative gridless approach is the Sliding Frank-Wolfe (SFW) algorithm \cite{denoyelle2019sliding}, which solves the Beurling LASSO problem, i.e., a traditional LASSO in the continuum \cite{Beurling2012exact}. SFW has been extended to 3D acoustic source localization in a grid-free setting \cite{chardon2021gridless}, and the choice of the regularization parameter is vital in obtaining accurate solutions.

In this work, we expand our preliminary work \cite{pandey2022parametric} on linear trajectories and grid-based algorithms to include general trajectories and perform estimation beyond the grid. Specifically, our contributions are:
\begin{itemize}
    \setlength\itemsep{0em}
    \item  We develop two trajectory models to account for dynamic source DOA: (a) bandlimited trajectory model and (b) polynomial trajectory model. 
    \item We develop two gridless algorithms to estimate the trajectory parameters: (a) SFW for trajectory localization (TL-SFW) and (b) NOMP for trajectory localization (TL-NOMP).
    \item We formulate wideband signal models and develop extensions of TL-SFW and TL-NOMP to perform trajectory localization using wideband signals.
    \item  We do a comprehensive performance analysis of proposed signal models and algorithms to study the impact of signal-to-noise ratio (SNR), number of snapshots, resolution limits, grid step-size, and computational complexity.
\end{itemize}

The structure of this paper is as follows: In Section \ref{sec:sigmodel}, we provide a brief overview of the conventional DOA signal model as well as the proposed trajectory model. In Section \ref{sec:gridbased_tlmodel}, we discuss the grid-based algorithms for trajectory localization. In section \ref{sec:gridless_Sigmod}, we describe two gridless techniques for trajectory localization, namely TL-SFW and TL-NOMP; their wideband extensions are also discussed. In Section \ref{sec:results_sim}, we present the simulation results, and Section \ref{sec:conclusions} concludes the paper.

In this paper, we denote the scalars, vectors, and matrices by lowercase, lowercase bold, and uppercase bold letters, respectively. $\X^H$ denotes the Hermitian matrix of $\X$. $||.||_\mathcal{F}$ stands for frobenious norm of a matrix and $|.|$ denotes the absolute value.
\section{\label{sec:sigmodel} Signal Model}
This section provides a brief overview of the static DOA model and proposes the parametric trajectory model. To model complex trajectories, polynomial and bandlimited models are used. The linear trajectory model, proposed in our earlier work \cite{pandey2022parametric,shreyas_TLDL}, can be recovered as a special case of the polynomial model.
\subsection{\label{subsec:static_doa} Static DOA}
In this subsection, the DOA is assumed to be constant within a block. Let $\y \in \C^{N}$ be the measurement vector received from an $N-$sensor uniform linear array (ULA), when $K$ sources are present:
\begin{align}
    \y &= \sum_{k=1}^{K}{\a(\theta_{k}) x_{k}} + \n = \A \x + \n \
\end{align} 
where $\A = [\a(\theta_1) \ldots \a(\theta_K)]$ is a matrix whose columns are steering vectors where $\a(\theta_{k})$ is steering vector corresponding to the source direction $\theta_{k}$ and $k = 1,\hdots, K$. Under the far-field assumption, the steering vector for $\theta_k$ direction is expressed as $\a(\theta_{k}) = [1, e^{j2\pi \frac{d \sin\theta_k}{\lambda}},\hdots, e^{j2\pi \frac{(N-1)d \sin\theta_k}{\lambda}}]$. $\x = [x_1, \ldots, x_K]$ is the source amplitude vector and $\n \in \C^{N}$ is the additive noise. $\lambda$ represents the wavelength of narrowband sources, and $d$ is the inter-sensor spacing in ULA. 

When a sequence of $L$ observations is available, the above narrowband model can be extended to multiple measurement vector (MMV) model~\cite{wipf2007,gerstoft2016} as:
\begin{align}
 \textbf{Y} = \textbf{AX} + \textbf{N}
    = [\A\x_{1} \ldots \A\x_{L}] + \textbf{N} \,
  \label{eq:prev_sigmodel}
\end{align}
where $\Y = [\y_1 \hdots ~\y_L] \in \C^{N\times L}$ is the $L$ snapshot observation matrix, $\X = [\x_1 \hdots \x_L] \in \C^{K\times L}$ represents the source amplitudes of $K$ sources over $L$ snapshots, and $\textbf{N} = [\n_1 \hdots \n_L] \in \C^{N\times L}$ accounts for the additive noise across $L$ snapshots. Under static DOA assumption, the source directions $(\theta_k)$ do not change with time and are determined by analyzing the block of $L$ snapshots.
\subsection{\label{subsec:TL_model} Parametric models for DOA trajectory}
In practical situations, sources are often in motion, making the assumption of constant DOA impractical. This presents a challenge in accurately estimating the DOA for moving sources. To overcome this issue, in our previous work \cite{pandey2022parametric}, we modelled and estimated linear DOA trajectories within block duration. However, the linear assumption does not always hold true, as sources can exhibit complex, non-linear motion. To address this limitation, we introduce two general trajectory models that can capture linear as well as non-linear motion -- polynomial trajectories and bandlimited trajectories.
\subsection{\label{subsec:polynomial_model} Polynomial trajectory model}
We define a $p^{\text{th}}$ order polynomial trajectory as a function of snapshot number as
    \begin{align}
\theta^{l} =~ &\phi + \sum_{p=1}^P \alpha_p \left({\frac{l}{L-1}}\right ) ^p
    \label{eq:polynomial_model}
    \end{align}
where $\theta^{l}$ represents the DOA at $l^{\text{th}}$ snapshot and $\boldsymbol{\omega} = (\phi,\alpha_1,\hdots,\alpha_p)$ denotes the vector of trajectory parameters for a source. The first order polynomial $(p=1)$ corresponds to the linear trajectory \cite{pandey2022parametric} model,
\begin{align}
    \theta^{l} = \phi +  \alpha_1 \left(\frac{l}{L-1}\right), \quad l = 0,1,\ldots,L-1
 \label{eq:linear_doa}
\end{align}
 whereas the zeroth order polynomial $(p=0)$ corresponds to the static DOA case. Note that increasing the number of parameters in the model allows for complex trajectories but, at the same time, leads to higher computations in the trajectory estimation algorithms.
\subsection{\label{subsec:bandlimited_model} Bandlimited trajectory model}
 Alternately, we can use the bandlimited model as discussed in \cite{tabaghi2019kinetic} to generate trajectories,
 \begin{align}
    \theta^{l} = \phi + \sum_{q=1}^{Q} \left\{ \alpha_q \sin{q\nu l} + \beta_q \cos{q\nu l} \right \} 
    \label{eq:bandlimited_doa}
    \end{align}  
where $\nu$ denotes the fundamental frequency of sinusoidal signals to be added and $\boldsymbol{\omega} = (\phi,\alpha_1,\hdots,\alpha_Q,\beta_1,\hdots,\beta_Q)$ denotes the vector of trajectory parameters for a source. These trajectories are guaranteed to be bandlimited, with the maximum frequency being $Q \nu$. We choose $Q$ based on expected DOA changes within a block. As in the case of polynomial trajectories, increasing $Q$ increases the computational cost of trajectory estimation algorithms.
\subsection{\label{subsec:observation_model}Observation model}
Let $\boldsymbol{\omega}_k \in \boldsymbol{\Psi}$ be the vector of parameters defining the $k^{\text{th}}$ source DOA trajectory, where $\boldsymbol{\Psi}$ is the continuous trajectory space. Define $\tilde{\A}(\boldsymbol{\omega}_k) \in \C^{N \times L}$ to be the trajectory steering matrix containing all the steering vectors as the DOA varies for the $k^{\text{th}}$ trajectory, i.e., $\tilde{\A}(\boldsymbol{\omega}_k) = \left [\a(\theta_k^{1}) \ldots \a(\theta_k^{L}))\right] = \left [\a_{1}^k \ldots \a_{L}^k\right] $, where $\theta^l$ represents the $l^{\text{th}}$ snapshot DOA in an $L$-length block. Let $\tilde{\X}_{k} = \text{diag}(\x^{k}), \x^{k} = [x^{1}_{k} \ldots x^{L}_{k}]^{T}$ be the diagonal matrix of $L$ complex amplitudes for the $k^{\text{th}}$ source. Thus, the MMV observation matrix when $K$ sources are present can be expressed as,
\begin{align}
    \Y &= \sum_{k=1}^{K}{\tilde{\A}(\boldsymbol{\omega}_k)} \tilde{\X}_{k} + \textbf{N}  
    = \sum_{k=1}^{K}{\tilde{\A}_{k} \tilde{\X}_{k}} + \textbf{N} \,, \\
    \Y &= \Bar{\A}(\mathcal{W}) \Bar{\X} + \textbf{N} = \Bar{\A} \Bar{\X} + \textbf{N} \,,
    \label{eq:newsigmodel}
\end{align}
where $\Bar{\X} = [\tilde{\X}_{1} \hdots \tilde{\X}_{K}]^{T}$, $\Bar{\A}(\mathcal{W}) = [\tilde{\A}_{1} \hdots \tilde{\A}_{K}]$, and $\mathcal{W} = \{\boldsymbol{\omega}_1,\hdots, \boldsymbol{\omega}_K\} \subset \boldsymbol{\Psi}$. Here $\Bar{\X}$ consists of $K$ diagonal matrices (of size $L \times L$) stacked vertically. Let $\mathcal{X}^{L}_{K}$ be the set of all such vertically stacked diagonal matrices, thus $\Bar{\X} \in \mathcal{X}^{L}_{K}$.

In contrast to the static DOA MMV model \eqref{eq:prev_sigmodel}, \eqref{eq:newsigmodel} represents the dynamic DOA MMV model, which accounts for source motion. In trajectory localization, our aim is to estimate parameters $(\boldsymbol{\omega}_k)$ defining the trajectory for all the sources from the given observation matrix.
\section{\label{sec:gridbased_tlmodel} Grid-based algorithms}
Grid-based algorithms use a predefined grid where each grid point represents a possible trajectory parameter to be estimated. The algorithm then analyzes the array measurements to determine the most likely parameters by comparing the signal characteristics at different grid points. In this section, we discuss grid-based methods for trajectory localization. We briefly describe existing methods \cite{pandey2022parametric} of TL-CBF and TL-SBL and introduce an extension of orthogonal matching pursuit for the trajectory model called TL-OMP. We conclude this section by showcasing grid-based TL algorithms for linear trajectory estimation with $\boldsymbol{\omega} = (\phi, \alpha)$ as described in \eqref{eq:linear_doa}.
\subsection{\label{subsec:TL_CBF}TL-CBF}
\begin{figure*}
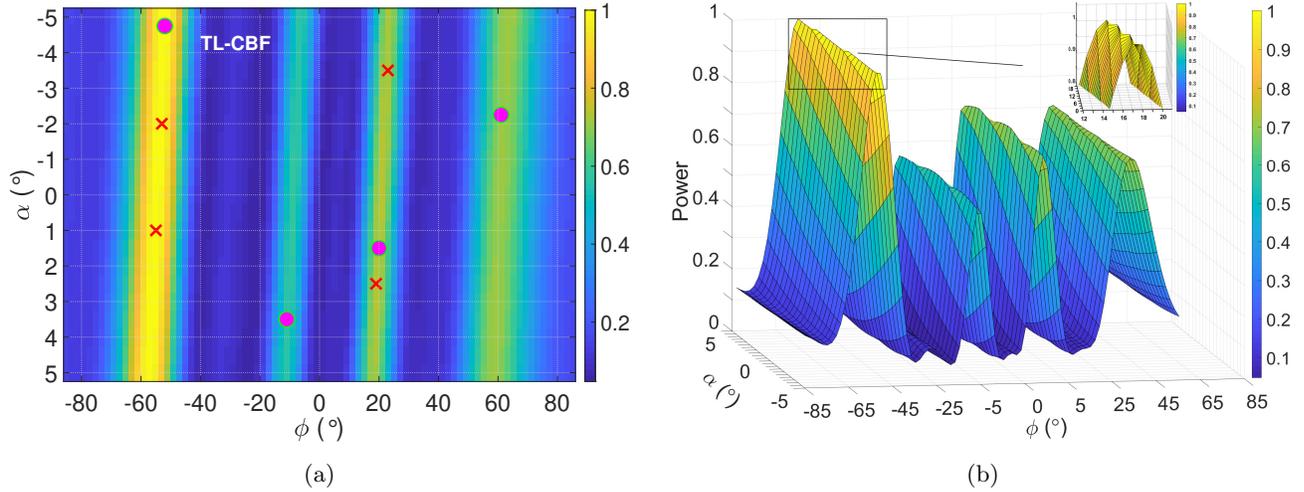

    \figline{
        \fig{TL_CBF_spectrum_final.eps}{.46\textwidth}{(a)}\label{fig:tlcbf1}
        \fig{3d_plot_final.eps}{.46\textwidth}{(b)}\label{fig:tlcbf2}
        }
    \caption{{TL-CBF spectrum for $4$ source trajectories with both on-grid and off-grid parameters. \ref{fig:tlcbf1}: 2D spectrum with true parameters $(-11,3.5)$, $(20,1.5)$, $(61, -2.25)$ and $(-52,-4.75)$ [shown by circle], detected and assigned peaks are shown by red cross and \ref{fig:tlcbf2}: 3D plot with inset showing spurious peaks around a single source.}}
    \vspace{-0.2cm}
\end{figure*}

A modification of the conventional beamforming (CBF) \cite{van1988beamforming} algorithm for the linear trajectory model is presented in \cite{pandey2022parametric}. We refer to it as trajectory localization-based CBF, i.e. TL-CBF. The original CBF algorithm computes the angular power spectrum at a predefined DOA grid by analyzing the correlation between the observations and the steering vectors \cite{compressivebeamforming}. The DOA estimates are determined from the peaks of this angular power spectrum. The TL-CBF extends this by computing the power spectrum using the following expression,
\begin{align}
    P_{\text{TL-CBF}}(\boldsymbol{\omega}) =  \frac{1}{L} \sum_{l=1}^{L} |\a^H_l(\boldsymbol{\omega})\,\y_l|^2
    \,,
    \label{eq:cbf_spec}
\end{align}
where the power spectrum $ P_{\text{TL-CBF}}(\boldsymbol{\omega})$ is two dimensional. The power is computed over a discrete trajectory space $(\boldsymbol{\omega} \in \boldsymbol{\Psi}_d)$ with $M$ potential grid points for $\boldsymbol{\omega}$. The locations of peaks in the spectrum are the estimated DOA trajectories. Figure \ref{fig:tlcbf1} and \ref{fig:tlcbf2} shows the 2D and 3D TL-CBF spectrum \eqref{eq:cbf_spec}, and the locations of the peaks provide trajectory parameters. 
\begin{figure*}
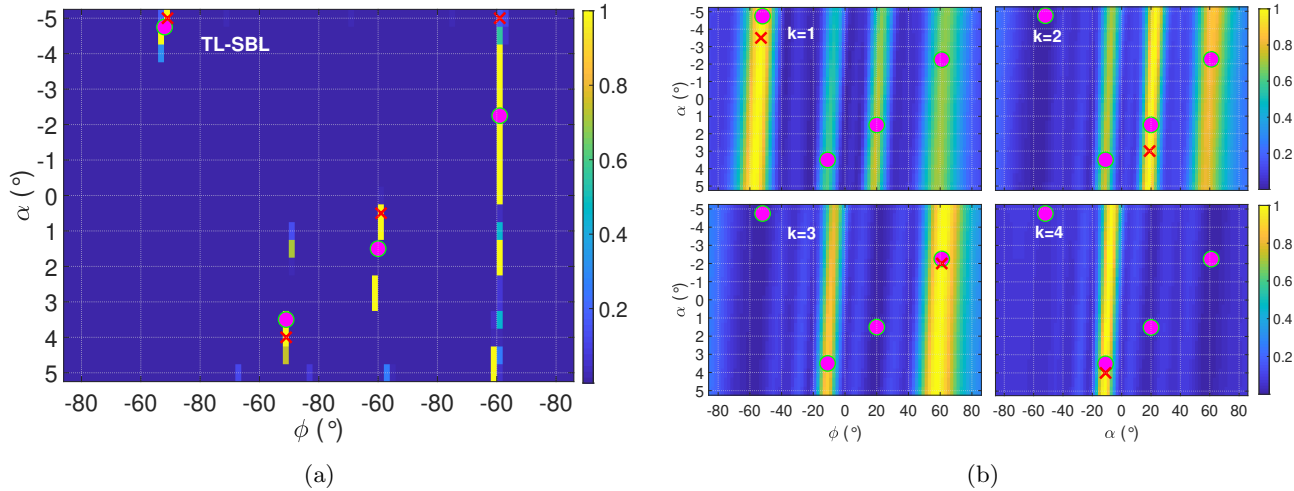

    \figline{
        \fig{TL_SBL_spectrum_final.eps}{.46\textwidth}{(a)}\label{fig:tlsbl}
        \fig{OMP_spectrum_final.eps}{.46\textwidth}{(b)}\label{fig:tlomp}
        }
    \caption{{Spectrum obtained for $4$ source trajectories with true parameters $(-11,3.5)$, $(20,1.5)$, $(61, -2.25)$ and $(-52,-4.75)$ [circle]. Detected and assigned peaks are shown by red cross. \ref{fig:tlsbl}: TL-SBL spectrum and \ref{fig:tlomp}: TL-OMP spectrum at each iteration.}}
   \vspace{-0.5cm}
\end{figure*}
\subsection{\label{subsec:TL_SBL}TL-SBL}
Sparse Bayesian learning (SBL) is a well-known compressive sensing method \cite{multifreq2019sparse, gerstoft2016}. A derivative of this method, TL-SBL, has been developed and applied to estimate DOA trajectory parameters \cite{pandey2022parametric}. The TL-SBL method is based on a sparse modelling framework, and the update rule for computing the TL-SBL spectrum is given as
 \begin{align}
    \hat \gamma_m^{\text{new}} = \hat \gamma_m^{\text{old}} \, \frac{\y_{v}^H \bSigma_{\y_{v}} \hat{\A}_m \hat{\A}_m^H \bSigma_{\y_{v}}^{-1} \y_{v}}{\text{Tr}[\bSigma_{\y_{v}}^{-1} \hat{\A}_m \hat{\A}_m^H] } \,,
      \label{eq:sbl_spec_final}
 \end{align}
where $\hat{\A}_m = \I_{L} \otimes \tilde{\A}_{m}$, and $\text{Tr}[\cdot]$ denotes trace of a matrix. The $m^\text{th}$ grid point represents a potential source ($\boldsymbol{\omega}_m$) with corresponding to the trajectory steering matrix $\tilde{\A}_{m}$. The vector $\bgamma = [\gamma_1, \hdots, \gamma_m]$ denotes the variance of source amplitude and, due to the hierarchical property of SBL, turned out to be sparse. The locations of non-zero entries of $\bgamma$ signify the source DOA trajectory estimates. An illustration of the TL-SBL spectrum \eqref{eq:sbl_spec_final} is shown in Figure \ref{fig:tlsbl} and features well-defined peaks. 
\subsection{\label{subsec:TL_OMP}TL-OMP}
We modify the orthogonal matching pursuit (OMP) algorithm \cite{mallat1993matching,cai2011orthogonal} to estimate the trajectory parameters. This greedy method iteratively selects the atoms from the dictionary on which the projection of the residual measurement matrix is maximum,
\begin{align}
      \hat{\boldsymbol{\omega}} = \underset{\boldsymbol{\omega} \in \boldsymbol{\Psi}_d} {\text{arg\,max}}~ \frac{1}{L} \sum_{l=1}^{L} \left|\textbf{a}^H_l\left(\boldsymbol{\omega}\right) \, \r_l^{[k-1]}\right|^2
 \label{eq:tlomp_obj}     
\end{align}
where $\r_l^{[k-1]}$ represents the residual at $l^\text{th}$ snapshot for $k^{\text{th}}$ iteration. The residual for the next iteration is,
\begin{align}
    \r_l^{[k]} = \r_l^{[k-1]} - \P_l ~\r_l^{[k-1]}
    \label{eq:residual}
\end{align}
where $ \P_l = \a_l*(\a_l^H*\a_l)^{-1} \a_l^H$ is the projection matrix. This ensures that the residual observation vectors (at each of the $l^\text{th}$ snapshots) are orthogonal to the corresponding steering vectors of the estimated source trajectories. The residual is initialized to the observation vector $\r_l^{[0]} = \y_l$.
TL-OMP is a greedy algorithm as it makes locally optimal choices at each step without considering the global impact, leading to suboptimal solutions. The TL-OMP spectrum \eqref{eq:tlomp_obj} at various iterations are shown in Figure \ref{fig:tlomp}. At each iteration, a source is found, and the residual is computed for the next iteration.
\subsection{\label{subsec:gridbased_example}Example}
We compare the grid-based algorithms for linear DOA trajectories. Figure \ref{fig:tlcbf1}, \ref{fig:tlsbl}, and \ref{fig:tlomp} show the 2D spectrum obtained from TL-CBF, TL-SBL and TL-OMP, respectively. Observations are generated using a 10-sensor ULA with $\frac{\lambda}{2}$ spacing. In each block, $L = 30$ snapshots are processed at $5$ dB SNR. The grid over linear parameters are set as ${\phi} = \{-85:2:85\}$ and ${\alpha} \in \{-5:0.5:5\}$. Four sources are present with trajectories $\{(-11,3.5), (20,1.5), (61,-2.25), (-52,-4.75)\}$. These include both on-grid and off-grid sources. Figures indicate both true and estimated trajectory parameters.

It can be seen from Figure \ref{fig:tlcbf1} and \ref{fig:tlcbf2} that TL-CBF has broad peaks, which makes it incapable of discerning closely spaced trajectories leading to poor resolution. In addition, there are numerous spurious peaks associated with each source (see Figure \ref{fig:tlcbf2} inset), which can cause repeated detection of the same source. 

In contrast, the TL-SBL spectrum offers higher resolution than TL-CBF but is computationally intensive as the size of the search grid increases, making it unsuitable for real-time applications. On the other hand, TL-OMP can estimate the trajectory parameters accurately, but it is a greedy algorithm. The grid-based algorithms are prone to bias errors when the parameters are off-grid. In this section, we only discussed the case of linear trajectories, but these algorithms can be extended to other trajectories with the corresponding results presented later.
\section{Gridless trajectory localization\label{sec:gridless_Sigmod}}
The performance of grid-based algorithms is limited when the true DOAs deviate from the grid or when the grid is too coarse, resulting in low resolution. Additionally, finer grid results in increased computational cost. To address this limitation, we describe an alternate model for \eqref{eq:newsigmodel} and formulate the Beurling LASSO problem for gridless trajectory localization. To solve this, we propose the TL-SFW and TL-NOMP algorithms and extend them for wideband signals as well.
\subsection{\label{subsec:lasso_formulation}Beurling LASSO}
Let there be $K$ sources with the trajectory parameters $\mathcal{W} = \{\boldsymbol{\omega}_1,\hdots, \boldsymbol{\omega}_K\} \subset \boldsymbol{\Psi}$ and diagonal matrices $\tilde{\X}_k$ consisting of amplitudes of the $L$ snapshots. Using Dirac mass $\delta_{\boldsymbol{\omega}}$ to represent a source with trajectory parameter $\boldsymbol{\omega} \in \boldsymbol{\Psi}$, we can reformulate \eqref{eq:newsigmodel} as,
\begin{align}
    \textbf{Y} &= \int_\Psi \tilde{\A}(\boldsymbol{\omega}) \, d\boldsymbol{\Delta} + \textbf{N} \,,
    \label{eq:gridless_model} \\
    \boldsymbol{\Delta} &= \sum_{k=1}^{K} \tilde{\X}_k \, \delta_{\boldsymbol{\omega}_k} \,,
    \label{eq:measure}
\end{align}
where $\boldsymbol{\Delta}$ is the measure representing all the sources.
A Beurling LASSO problem can now be constructed as,
\begin{align}
\boldsymbol{\Delta}^{*} &= 
    \argmin \limits_{\boldsymbol{\Delta} \in \mathcal{M}} \, \frac{1}{2} 
    \left|\left| \int_\Psi \tilde{\A}(\boldsymbol{\omega}) \, d\boldsymbol{\Delta} - \textbf{Y} \right|\right|_{\mathcal{F}}^{2} 
    + \lambda |\boldsymbol{\Delta}| \,,
\label{Beurling_LASSO}
\end{align}
where $\mathcal{M}$ is the set of complex measures defined on $\boldsymbol{\Psi}$, $\lambda$ is the regularization parameter, and $|\boldsymbol{\Delta}|$ represents any sparsity inducing norm of the measure $\boldsymbol{\Delta}$. The regularization parameter $\lambda$ can be tuned to find the number of sources. In this work, we assume the number of trajectories to be known; thus, we set $\lambda = 0$ and develop greedy iterative algorithms \cite{chardon2021gridless}. From the solution $\boldsymbol{\Delta}^{*}$, we obtain estimates for the trajectory parameters $\mathcal{W}$ and their corresponding amplitudes using \eqref{eq:measure}. In presence of wideband observations $\textbf{Y}_f, f = 1,2,\ldots,F$, a multi-frequency Beurling LASSO can be constructed as,
\begin{align}
\boldsymbol{\Delta}^{*} &= 
    \argmin \limits_{\boldsymbol{\Delta} \in \mathcal{M}} \, \frac{1}{2} 
    \sum_{f=1}^F \left|\left| \int_\Psi \tilde{\A}_f(\boldsymbol{\omega}) \, d\boldsymbol{\Delta} - \textbf{Y}_f \right|\right|_{\mathcal{F}}^{2} \,.
\label{wideband_sfw}
\end{align}
\subsection{\label{subsec:TL-SFW} Sliding Frank-Wolfe algorithm (TL-SFW)}
We solve the Beurling LASSO problem \eqref{Beurling_LASSO} using greedy ($\lambda = 0$) Sliding Frank-Wolfe (SFW) algorithm \cite{Beurling2012exact,denoyelle2019sliding,chardon2021gridless}. The SFW algorithm for trajectory localization (TL-SFW) is detailed in Algorithm \ref{algorithm:TL-SFW}. We iteratively solve \eqref{Beurling_LASSO} by adding one source at a time. An empty set is denoted as $\emptyset$.

$\R^{[k]}$ denotes the $N \times L$ residual matrix at the end of iteration $k$ and is initialized as $\R^{[0]} = \Y$.
Each iteration over $K$ trajectories consists of the following steps:
\begin{enumerate}[label=(\roman*)]
\setlength\itemsep{0em}
\item {\bf{Add a source:}} Solve \eqref{eq:tlomp_obj} to find a coarse trajectory estimate on the predefined grid $\boldsymbol{\Psi}_d$. Use this estimate as initialization to solve the global optimization problem $(a)$ in Algorithm \ref{algorithm:TL-SFW} to obtain $\boldsymbol{\omega}^*$.
\item {\bf{Amplitude estimation:}} Initialize all the $k$ source amplitudes as $\text{diag}(\tilde{\X}_{k}) = \text{diag}(\tilde{\A}^H(\boldsymbol{\omega}_k) \Y)$ using the estimated trajectory parameters $\mathcal{W}^{[\frac{k-1}{2}]}$. Solve $(b)$ to obtain optimized amplitudes $\Bar{\X}^{[\frac{k-1}{2}]}$.
\item {\bf{Joint estimation:}} Jointly optimize the trajectory parameters and amplitudes by solving $(c)$. Initialization is done using $\mathcal{W}^{[\frac{k-1}{2}]}$ and $\Bar{\textbf{X}}^{[\frac{k-1}{2}]}$ for this non-convex optimization problem.
\end{enumerate}

The algorithm is proven to converge in a finite number of iterations under certain constraints \cite{denoyelle2019sliding}. Optimizations $(a)$, $(b)$, and $(c)$ are performed using the sequential quadratic programming algorithm \cite{nocedal1999numerical} in the MATLAB 2018b function \texttt{fmincon}. For wideband observations, problems $(a)$ and $(c)$ are respectively modified as,
\begin{align}
    \boldsymbol{\omega}^* = \underset{\boldsymbol{\omega} \in \boldsymbol{\Psi}} {\text{argmax}} & \, \frac{1}{L} \, \sum_{f=1}^{F} \sum_{l=1}^{L} \left|\textbf{a}^H_{lf}(\boldsymbol{\omega})\, \r_{lf}^{[k-1]}\right|^2
    \label{wideband_postion_opt} \\ 
    \{\Bar{\textbf{X}}^{[k]}_{f}\}_{f=1}^{F}, \mathcal{W}^{[k]} &= 
    \underset{\mathcal{W} \subset \boldsymbol{\Psi},\Bar{\X}_f \in \mathcal{X}^{L}_{k}}{\text{argmin}} \, \frac{1}{2} \, \sum_{f=1}^{F} \, \left|\left| \Bar{\textbf{A}}_f(\mathcal{W}) \, \Bar{\textbf{X}}_f  - \textbf{Y}_f \right|\right|_\mathcal{F}^2 \,.
    \nonumber
\end{align}
For wideband processing, the trajectory parameters are estimated using the averaged spectrum over $F$ frequencies. The optimization $(b)$ is solved $F$ times to obtain the amplitudes $\Bar{\textbf{X}}_f$ at each frequency. As the number of frequencies increases, the number of unknown parameters also increases, leading to a higher computational cost.
\vspace{0.7cm}
\begin{algorithm}[t]
\caption{TL-SFW pseudo-code to solve \eqref{Beurling_LASSO}}
\label{algorithm:TL-SFW}
\begin{algorithmic}
\STATE {1. $\mathcal{W}^{[0]} \leftarrow \emptyset, \R^{[0]} \leftarrow \textbf{Y}, tol = 1e^{-10} $} \vspace{4pt}
\STATE{2. \textbf{for} $k = 1,\hdots,K$}  \vspace{4pt}
\STATE{3. $\quad$ \underline{Find the next source:} }\vspace{4pt}
\STATE{$\quad\quad$ $~ \boldsymbol{\omega}^* = \underset{\boldsymbol{\omega} \in \boldsymbol{\Psi}} {\text{arg\,max}}~ \frac{1}{L} \sum_{l=1}^{L} \left|\textbf{a}^H_l(\boldsymbol{\omega})~\r_l^{[k-1]}\right|^2~~~~~~~(a)$} \vspace{4pt}
\STATE{4. $\quad$ $\mathcal{W}^{[\frac{k-1}{2}]} = \{ \mathcal{W}^{[k-1]},~ \boldsymbol{\omega}^* \} $}\vspace{4pt}
\STATE{5. $\quad$ \underline{Optimize the amplitude:} } \vspace{4pt}
\STATE{$\quad\quad$ $\Bar{\X}^{[\frac{k-1}{2}]} = \underset{\Bar{\X} \, \in \, \mathcal{X}^{L}_{k}}{\text{arg\,min}} \, \frac{1}{2} ~\left|\left|\Bar{\textbf{A}}(\mathcal{W}^{[\frac{k-1}{2}]}) \, \Bar{\textbf{X}} - \textbf{Y}\right|\right|_\mathcal{F}^2~~~~~(b)$}\vspace{4pt}
\STATE{6. $\quad$ \underline{Optimize the amplitudes and parameters:}} \vspace{4pt}
\STATE{$\quad\quad$ $\Bar{\textbf{X}}^{[k]}, \mathcal{W}^{[k]}   = \underset{\mathcal{W} \subset \boldsymbol{\Psi},\Bar{\X} \in \mathcal{X}^{L}_{k}}{\text{arg\,min}} \, \frac{1}{2} \, \left|\left| \Bar{\textbf{A}}(\mathcal{W}) \, \Bar{\textbf{X}}  - \textbf{Y} \right|\right|_\mathcal{F}^2~~~(c)$}\vspace{4pt}
\STATE{7. $\quad$ $\textbf{R}^{[k]} \leftarrow \textbf{Y}-\Bar{\textbf{A}}(\mathcal{W}^{[k]}) \, \Bar{\textbf{X}}^{[k]} $} \vspace{4pt}
\STATE{8. \textbf{end for}} \vspace{5pt}
\end{algorithmic} \vspace{5pt}

{MATLAB \texttt{fmincon} is used to solve equations (a), (b), (c)}\\[4pt] \hrule \vspace{2pt} \hrule
\end{algorithm}
\subsection{\label{subsec:TL-NOMP}Newtonized OMP (TL-NOMP)}
Newtonized orthogonal matching pursuit (NOMP) is a variant of OMP that incorporates Newton refinements to obtain precise off-grid estimates \cite{mamandipoor2016newtonized,yang2020two}. The NOMP algorithm for trajectory localization (TL-NOMP) is given in Algorithm \ref{algorithm:TL-NOMP}. NOMP has three main steps when adding a new source:
\begin{enumerate}[label=(\roman*)]
    \item \textbf{Find a source:} Obtain an initial coarse estimate $\boldsymbol{\omega}^*$ of source trajectory parameter by searching over the grid $\boldsymbol{\Psi}_d$ using \eqref{eq:tlomp_obj} and estimate the corresponding amplitudes $\tilde{\X}^*$.
    \item \textbf{Local Newton refinement:} Compute the Hessian matrix $(\H)$ and gradient vector $(\boldsymbol{g})$ for the objective in \eqref{Beurling_LASSO}. Refine the on-grid trajectory parameter estimate using single-step Newton's method over the continuum $\boldsymbol{\Psi}$.
    \item \textbf{Global cyclic refinement:} Starting with the current residual $\R^*$ as the observation, add back each of the identified sources (one at a time) and optimize parameters using Local Newton refinement. Repeat until the convergence criteria is met.
\end{enumerate}
The local Newton refinement provides an improvement on the initial on-grid parameter estimate, whereas the global cyclic refinement provides a feedback mechanism to improve the estimates accumulated from previous iterations. At the end of the $k$-th iteration, the residual $\R^{[k]}$ is updated using \eqref{eq:residual} where data is orthogonally projected onto steering vectors corresponding to identified source trajectories. For the wideband implementation of NOMP, the objective in \eqref{wideband_sfw} is used instead.
\vspace{0.4cm}
\begin{algorithm}[t]
\caption{TL-NOMP pseudo-code to solve \eqref{Beurling_LASSO}}
\label{algorithm:TL-NOMP}
\begin{algorithmic}
\STATE {1.$\mathcal{W}^{[0]} \leftarrow \emptyset, ~\R^{[0]} \leftarrow \Y, ~tol = 1e^{-6} $} \vspace{4pt}
\STATE{2. \textbf{for} $k = 1,\hdots,K$}  \vspace{4pt}
\STATE{3. $\quad$ \underline{Find the next source:} }\vspace{4pt}
\STATE{$\quad\quad$ $~ \boldsymbol{\omega}^* = \underset{\boldsymbol{\omega} \in \boldsymbol{\Psi}_d} {\text{arg\,max}}~ \frac{1}{L} \sum_{l=1}^{L} \left|\textbf{a}^H_l(\boldsymbol{\omega})~\r_l^{[k-1]}\right|^2$} \vspace{4pt}
\STATE{$\quad$ $\quad$ $\text{diag}(\tilde{\X}^*) = \text{diag}\big(\tilde{\A}^H(\boldsymbol{\omega}^*) \R^{[k-1]}\big)$ }\vspace{4pt}
\STATE{4.$\quad$ \underline{Local Newton refinement:}}\vspace{4pt}
\STATE{$\quad \quad$ $\boldsymbol{\omega}^{*} = \boldsymbol{\omega}^{*} - \H^{-1}\boldsymbol{g} $\quad$ $} \vspace{4pt}
\STATE{$\quad \quad$ $\text{diag}(\tilde{\X}^*) = \text{diag}\big(\tilde{\A}^H(\boldsymbol{\omega}^*) \R^{[k-1]}\big)$ }\vspace{4pt}
\STATE{5. $\quad$ $\mathcal{W}^{[\frac{k-1}{2}]} = \{ \mathcal{W}^{[k-1]},~ \boldsymbol{\omega}^* \} $}\vspace{4pt}
\STATE{6. $\quad$ \underline{Global cyclic refinement:} } \vspace{4pt}
\STATE{$\quad \quad \quad$ $\textbf{R}^* \leftarrow \textbf{Y}-\Bar{\textbf{A}}(\mathcal{W}^{[\frac{k-1}{2}]}) \, \Bar{\textbf{X}}^{[\frac{k-1}{2}]} $} \vspace{4pt}
\STATE{$\quad \quad \quad$ \textbf{while} $\left| ||\R^{[k-1]}||_{f}^2 - ||\R^*||_{f}^2 \right | < tol$} \vspace{4pt}
\STATE{$\quad \quad \quad \quad$\textbf{for} $i = 1,\hdots,k$}  \vspace{4pt}
\STATE{$\quad \quad \quad \quad \quad$ $\hat{\R} = \R^* + \tilde{\A}(\boldsymbol{\omega}_i) \tilde{\X}_i$} \vspace{4pt}
\STATE{$\quad \quad \quad \quad \quad$ $\text{diag}(\tilde{\X}_i) = \text{diag}\big(\tilde{\A}^H(\boldsymbol{\omega}_i) \hat{\R}\big)$ }\vspace{4pt}
\STATE{$\quad \quad \quad \quad \quad$ Local Newton refinement of $\boldsymbol{\omega}_i$ and $\tilde{\X}_i$}\vspace{4pt}
\STATE{$\quad \quad \quad \quad \quad$ $\R^{[k-1]} \leftarrow\R^*$,  $\R^* \leftarrow \hat{\R} - \tilde{\A}(\boldsymbol{\omega}_i) \tilde{\X}_i$ }\vspace{4pt}
\STATE{$\quad \quad \quad \quad$\textbf{end for}}\vspace{4pt}
\STATE{$\quad \quad \quad$\textbf{end while}}\vspace{4pt}
\STATE{7. $\quad$ Use \eqref{eq:residual} to find the orthogonal residual $\R^{[k]}$} 
\STATE{8. \textbf{end for}} \vspace{5pt}
\end{algorithmic}
\end{algorithm}
 \begin{figure*}
\sidebysidefigures{RMSE_vs_SNR_final.eps}{Evaluation of TL-methods for linear trajectory localization for various SNR values. RMSE vs SNR (top) and $P_d$ vs SNR (bottom). \label{fig:snr}}
/{snapshots_new_corrected.eps}{Evaluation of TL-methods for linear trajectory localization for various snapshots processed within a block. RMSE vs Snapshots (top) and $P_d$ vs Snapshots (bottom).\label{fig:rmse_vs_snapshots}}
\vspace{-0.7cm}
\end{figure*}
\section{Simulations\label{sec:results_sim}}
\subsection{\label{subsec:sim_setup}Simulation setup}
We demonstrate various algorithms using simulations with linear and as well as non-linear trajectories. The performance of TL-SFW and TL-NOMP are compared with TL-CBF, TL-SBL and TL-OMP. A $10-$ sensor uniform linear array (ULA) with inter-sensor spacing $d = \frac{\lambda}{2}$ is used. Unless stated otherwise, simulations are for linear trajectories and narrow-band signals.

For grid-based methods TL-CBF, TL-SBL and TL-OMP, we construct the following grid over trajectory parameters: $\phi \in \{-85:2:85\}$ and $\alpha \in \{-5:0.5:5\}$ resulting in a dictionary with $M = 86 \times 21 = 1806$ trajectory steering matrices $\tilde{\A}$.
Throughout the simulations we consider $L = 30$ snapshots within a block at an SNR of $5$ dB. The source amplitudes and noise are complex Gaussian of the form $a+jb$ where $a$ and $b$ are generated using zero-mean Gaussians. The variance of signal and noise are $\sigma_x^2$ and $\sigma_n^2$, respectively. The signal-to-noise ratio is defined as $\text{SNR} = 10 \text{log}_{10} (\frac{\sigma_x^2}{\sigma_n^2})$.
For TL-SBL, the noise variance is assumed to be known and directly used in the update rule. However, an update rule for estimating the noise variance can also be derived \cite{gerstoft2016,multifreq2019sparse,bohme1985source}.

To compare the localization accuracy of TL methods, we report root mean square error (RMSE). Let ${\theta}_k^{l}$ and $\hat{\theta}_k^{l}$ 
be the ground truth and estimated DOA obtained from trajectory parameters corresponding to the $k^{\text{th}}$ source. The RMSE for $k^{\text{th}}$ source is given by,
 \begin{align}
    \text{RMSE}_k = \sqrt{\frac{\sum_{l=0}^{L-1} \,(\theta_{k}^{l}-\hat{\theta}_{k}^{l})^2}{L}}, \quad k=1,\ldots, K.
    \label{eq:rmse_metric}
 \end{align}
We perform 100 Monte Carlo trials and report the RMSE averaged across all the trials and sources. 

For TL-CBF and TL-SBL, if $K$ sources are present, we identify $\hat{K} = K+2$ peaks in the power spectrum. By considering more peaks, we overcome the problem of spurious peaks and get the best possible estimates closer to true trajectories. 
The Optimal SubPattern Assignment (OSPA) \cite{schuhmacher2008consistent,ospa_2} is used to solve the assignment problem between the $\hat{K}$ estimated trajectories and $K$ true trajectories. 
Let $\hat{\mathcal{T}} \triangleq \left\{ \hat{T}_1,\hdots,\hat{T}_{\hat{K}} \right\} $ be the set of $\hat{K}$ estimated trajectories and $\mathcal{T} \triangleq \left\{ T_1,\hdots,T_K \right\} $ be the set of $K$ true trajectories. The OSPA metric for sets $\mathcal{T}$ and $\hat{\mathcal{T}}$ is defined as:
 \begin{equation}
     \text{OSPA}(\mathcal{T},\hat{\mathcal{T}}) \triangleq \left[ \frac{1}{\hat{K}} \underset{\pi \in \Pi_{\hat{K}}}{\min} \sum_{k=1}^K d_c({T}_{k},\hat{{T}}_{\pi(k)})^p + (\hat{K}-K)c^p \right]^\frac{1}{p} 
     \label{eq:ospa} 
 \end{equation}
where $K\leq \hat{K}$, the order parameter is $1 \leq p \leq \infty$  and $c$ is the cutoff parameter. $\Pi_{\hat{K}}$ denotes the set of all permutations of length $K$ with elements $\{ 1, \hdots \hat{K}\}$. The $d_c({T}_{k},\hat{{T}}_{\pi(k)}) \triangleq \min(c,d_t({T}_{k},\hat{{T}}_{\pi(k)})$, where $d_t({T}_{k},\hat{{T}}_{\pi(k)})$ denotes the error between two trajectories computed using \eqref{eq:rmse_metric}. We choose $p = 2$ and $c = 100$. 
 
Once assigned, a source is said to be detected if the RMSE between ground truth and the assigned track is less than the detection threshold of $5^\circ$. We report the probability of detection $P_d$, i.e. the percentage of detected sources. The average RMSE is reported only for detected sources. 
\begin{figure*}[t!]
\sidebysidefigures{phii_step_final.eps}{Error as function of parameter $\phi$ grid step-size with $L=30$ snapshots at 5dB SNR.\label{fig:phi_grid_step}}
/{RMSE_vs_separation_final.eps}{Error a function of source proximity ($\zeta$) with $L=30$ snapshots at 5dB SNR. RMSE vs $\zeta$ (top) and $P_d$ vs $\zeta$ (bottom).\label{fig:rmse_vs_separation}}
\vspace{-0.7cm}
\end{figure*}
\subsection{\label{subsec:sim_snr_linear}Signal-to-Noise ratio}
We perform simulations with SNR ranging from $-10$dB to $30$dB. Four source trajectories (linear) are processed in a block containing $L=30$ snapshots. The true trajectory parameters are $\mathcal{W} = \{(-11,3.5), (20,1.5), (61, -2.25), (-52,-4.75)\}$, such that some parameters are on-grid while the rest are off-grid. The minimum error achievable by on-grid methods for each of these trajectories are $0, 0.51, 0.15, \text{and}, 0.53$ respectively, giving an average of $0.30$. The error vs SNR and $P_d$ vs SNR plots are shown in Figure \ref{fig:snr}. 
At low SNR, TL-CBF has the lowest RMSE; however, it exhibits lower $P_d$ compared to other approaches as it fails to detect all the sources. Both TL-NOMP and TL-SFW outperform all the grid-based methods as they can optimize the parameters beyond the grid. As SNR increases, most algorithms reach saturation except TL-NOMP, which consistently enhances its performance. At low SNR, TL-SFW has a slightly better detection rate compared to TL-NOMP. TL-SBL error saturates to the value of $0.30$ beyond which its performance cannot improve since it can only find sources on the grid. It performs better than TL-OMP, which is a greedy algorithm. 



\subsection{\label{subsec:sim_snap_linear}Snapshots}
We evaluate algorithm performance with the number of snapshots ranging from 5 to 50 at 5 dB SNR. The true trajectory parameters are the same as above. Figure \ref{fig:rmse_vs_snapshots} shows that as the number of snapshots increases, the error decreases for all the algorithms. Both TL-SFW and TL-NOMP show superior performance compared to all the other methods. The grid-based methods exhibit higher error compared to grid-free methods due to the bias present while estimating off-grid trajectory parameters, which is regardless of the number of snapshots. TL-CBF has higher $P_d$ for fewer snapshots which reduces with increasing snapshot number. This is likely due to the presence of spurious peaks (Figure \ref{fig:tlcbf2}) which become more prominent with increasing snapshots \eqref{eq:cbf_spec}.


\subsection{\label{subsec:sim_gridstep}Grid step-size}
We analyze the impact of step-size $(\phi_\text{step})$ used for creating $\phi$ grid in trajectory localization tasks. The grid over $\alpha$ is fixed with $\alpha \in \{-5:0.5:5\}$ while the grid over $\phi$ is made coarser by increasing $\phi_\text{step}$ from $1$ to $10$.
Let $\boldsymbol{\phi}_g$ be the grid vector constructed using $\phi_\text{step}$ with $N_\phi$ grid points. For this $\phi_\text{step}$ experiment, the true parameters are $\left(\boldsymbol{\phi}_g(\lfloor N_\phi \times 0.2 \rfloor),3.5\right)$, $\left(\boldsymbol{\phi}_g(\lfloor N_\phi \times 0.45 \rfloor) + \frac{\phi_\text{step}}{2},1.5\right)$, $\left(\boldsymbol{\phi}_g(\lfloor N_\phi \times 0.65 \rfloor),-2.5\right)$ and $\left(\boldsymbol{\phi}_g(\lfloor N_\phi \times 0.9 \rfloor) + \frac{\phi_\text{step}}{2},-4.75\right)$ where $\lfloor. \rfloor$ denotes the floor of a real number. These source trajectories are chosen such that the true $\phi$ and $\alpha$ parameters have both on-grid and off-grid combinations. As the step-size increases, the grid becomes less refined, and the performance of grid-based methods is expected to degrade. Whereas TL-SFW and TL-NOMP are expected to perform better since they improve upon the initial on-grid estimates by performing optimization and refinement, respectively. This analysis is verified from simulation results shown in Figure \ref{fig:phi_grid_step}. The impact of grid step-size on gridless methods is low with TL-NOMP being most robust to coarseness of the $\phi$ grid. 

\begin{figure*}[t!]
\sidebysidefigures{quadratic_final.eps}{Quadratic model: True and estimated trajectories using TL-SFW and TL-NOMP for a single block at 5 dB SNR. The true trajectories are $(-40,-3,-1.4)$, $(-21,0.4,-3.6)$, $(10,-3.2,1.6)$, $(61,2.4,3.2)$.  \label{fig:multiple_tracks}}
/{Bandlimited_tracks.eps}{Bandlimited Model: True and estimated trajectories using TL-SFW and TL-NOMP for a single block at 5 dB SNR. The true trajectories are $(-60,-3.2,-4.6)$, $(-19,0.8,3)$, $(24,-1.5,-3.7)$, $(61,4.3,4)$  \label{fig:bandlimited_multiple_tracks}}
\vspace{-0.7cm}
\end{figure*}
%
\begin{figure*}
\sidebysidefigures{RMSE_vs_SNR_NL_Final.eps}{Performance of TL-methods for non-linear trajectory localization at various SNR values. RMSE vs SNR (top) and $P_d$ vs SNR (bottom).\label{fig:snr_nl}}
/{ROC_NL_SNR_final.eps}{Performance of TL-methods for non-linear trajectory localization with varying detection threshold at different SNR. \label{fig:roc_snr_nl}}
\vspace{-0.7cm}
\end{figure*}
\subsection{\label{subsec:sim_resolu}Resolution}
Resolution refers to the ability to distinguish between two nearby trajectories accurately. We consider $3$ sources with linear trajectory parameters as follows $\mathcal{W} = \{(0,3.5), (60,-4.5), (\zeta,2.5)\}$. The $3^{\text{rd}}$ source trajectory varies as we increase $\zeta$ from $-15$ to $15$. Specifically, its trajectory approaches that of the $1^{\text{st}}$ source and then diverges. We process $30$ snapshots at 5 dB SNR. The results are shown in Figure \ref{fig:rmse_vs_separation}. TL-CBF, TL-OMP and TL-SFW have low resolution when dealing with closely spaced trajectories, as indicated by the peaks in the RMSE plot. Both TL-SBL and TL-NOMP outperform other methods, with TL-NOMP having the lowest error among all the methods. The detection performance of TL-SBL is influenced by our approach of selecting five peaks from the spectrum and subsequently identifying the three closest tracks after source association. Though there is a dip in error for all algorithms around $\zeta \in [-3,3]$, it is likely due to repeated identification of the same source and cannot be attributed to superior resolution ability.


\subsection{\label{subsec:sim_nonlinear}Non-linear trajectories}
Sample non-linear trajectories, generated using $3$ parameter quadratic and bandlimited models, are shown in Figures \ref{fig:multiple_tracks} and \ref{fig:bandlimited_multiple_tracks}, respectively. Each trajectory spans over $L = 40$ snapshots. Estimated trajectories, by processing observations at $20$ dB SNR, using TL-SFW and TL-NOMP are shown as well. For both models, we construct the following grid over trajectory parameters: $\phi \in \{-85:2:85\}$ and $\alpha_1, ~\alpha_2,~ \beta_1 \in \{-5:0.5:5\}$, resulting in a dictionary with $M = 86 \times 21 \times 21 = 37926$ trajectory steering matrices $\tilde{\A}$. This is significantly larger than the number of grid points in the linear case.    

Figure \ref{fig:snr_nl} shows error vs SNR for non-linear trajectory estimation. For this simulation we set $L=30$ and use sources with polynomial trajectories: $\mathcal{W} = \{(-60,1,-3),~ (-31,0.4,3.6),~ (20,-3,2),~ (51,4,-0.2)\}$. TL-CBF frequently fails to detect trajectories giving a poor detection rate of $P_d \approx 40\%$. TL-NOMP performs worse than TL-OMP at low SNR (both in RMSE and $P_d$) but recovers at higher SNR values outperforming all other algorithms. TL-SFW shows marginal improvement over TL-OMP with its error saturating at high SNR.

All the results presented so far use a detection threshold of $5^{\circ}$. Here we investigate the effect of changing this detection threshold on detection probability $P_d$. Figure \ref{fig:roc_snr_nl} depicts the $P_d$ as the detection threshold is changed for select SNR values. As expected, an increase in the value detection threshold increases $P_d$. Similar to the inference from Figure \ref{fig:snr_nl}, at lower SNR performance of TL-OMP is better than that of TL-NOMP whereas TL-SFW shows superior detection performance at all SNR levels.

\begin{figure*}
\sidebysidefigures{complexity_snapshots.eps}{Complexity analysis of TL-methods for a varying number of snapshots. Linear trajectory model (top) and quadratic trajectory model (bottom).\label{fig:snapshots_complexity}}
/{wideband_SNR_all_final.eps}{Performance of wideband TL-methods for quadratic trajectories with different numbers of processed frequencies at various SNR.\label{fig:wideband_result}}
\vspace{-0.3cm}
\end{figure*}


\subsection{\label{subsec:complexity}Computational effort}
In this section, we present the computational time analysis of methods by varying snapshots from $5$ to $50$, at $5$ dB SNR. We conduct experiments on a desktop equipped with an Intel(R) Core(TM) i7-8700 CPU operating at 3.19 GHz $\times$ 6 cores and 32 GB of memory. Figure \ref{fig:snapshots_complexity} illustrates the computational time required by each method for estimating linear (top) and non-linear (bottom) trajectories. TL-CBF and TL-OMP exhibit high computational efficiency leading to significantly shorter execution times when compared to other methods. For non-linear trajectories, TL-SBL requires significantly longer execution times, even with a small number of snapshots. Hence, we omit TL-SBL results for the non-linear case. The computational requirements of TL-NOMP are higher than that of TL-SFW.


\begin{table*}[]
\centering
\begin{tabular}{|c|c|c|c|c|c|}
\hline
\textbf{Algorithm} & \begin{tabular}[c]{@{}l@{}}\textbf{Noise resilience}\end{tabular} & \textbf{Resolution} & \textbf{Effect of grid-step} & \textbf{Computation speed} & \begin{tabular}[c]{@{}l@{}}\textbf{Detection probability}\end{tabular}\\ \hline

\textbf{TL-CBF} & Low & Low & High & Fast & Low\\ \hline

\textbf{TL-SBL} & Medium & High & High & Slow & Medium\\ \hline

\textbf{TL-OMP} & High & Low & High & Fast & High \\ \hline

\textbf{TL-NOMP} & High & High & Low & Medium & Medium\\ \hline

\textbf{TL-SFW} & High & Medium & Medium & Fast & High\\ \hline
\end{tabular}
\caption{Comparative analysis of various algorithms for trajectory localization.}
\label{tab:summary}
\vspace{-0.3cm}
\end{table*}
\subsection{\label{subsec:sim_wideband}Wideband processing}
We generate wideband observations and apply TL algorithms. The TL-SFW processes the multi-frequency signals in a coherent manner \eqref{wideband_postion_opt}, whereas other TL methods process them non-coherently. We extend the TL-CBF and TL-OMP to wideband observations by summing the spectrum across frequencies in \eqref{eq:cbf_spec} and \eqref{eq:tlomp_obj}. We do not include wideband \cite{multifreq2019sparse,pandey2022experimental} TL-SBL due to its high computational complexity.
We examine the performance by increasing the number of frequencies processed as $F = 1, 3, 5, \text{and } 7$ with corresponding frequencies $1600$, $\{1400, 1600, 1800\}$, $\{1000, 1200, 1400, 1600, 1800\}$, and $\{1000, 1200, 1400, 1600, 1800, 2000, 2200\}$.
Figure \ref{fig:wideband_result} shows that as the number of frequencies increases, the performance improves.
The TL-NOMP shows the best performance among all and significantly improves over TL-OMP.
TL-SFW shows degraded performance when more frequencies are used, which could be due to the additional amplitude parameters it has to estimate as the number of frequencies increases.


\section{\label{sec:conclusions}Conclusion}
In this paper, we proposed two novel trajectory models: bandlimited and polynomial. We developed two gridless algorithms for localizing the DOA trajectories -- TL-SFW and TL-NOMP -- and demonstrated their superior performance in extensive simulations. We also extended the algorithms for wideband processing. Table \ref{tab:summary} provides a summary of the performance characteristics of various algorithms highlighting their noise resilience, resolution, sensitivity to grid-step, speed and detection probability. Among grid-based methods, TL-CBF and TL-OMP are fast but have low to moderate resolution, whereas TL-SBL is slow but has high resolution. Among the gridless methods, TL-SFW is preferable in scenarios where noise resilience, computational efficiency and detection rate are prioritized, while TL-NOMP is more suitable for applications that require noise resilience and high resolution and coarse parameter grids are tolerable. Overall, gridless algorithms outperform grid-based methods for trajectory localization.

\bibliography{sampbib}

\begin{thebibliography}{10}
\def\enquote#1,{``#1,''}
\def\enxquote#1{``#1''}
\expandafter\ifx\csname url\endcsname\relax
  \def\url#1{\texttt{#1}}\fi
\expandafter\ifx\csname urlprefix\endcsname\relax\def\urlprefix{URL }\fi
\providecommand{\bibinfo}[2]{#2}
\def\plainquote#1{``#1''}
\providecommand{\noopsort}[1]{}
\providecommand{\switchargs}[2]{#2#1}
\providecommand{\dourl}[1]{\href{http://#1}{\nolinkurl{#1}}}
  \def\eatspace #1{#1}

\bibitem{alenljung2019user}
\bibinfo{author}{B.~Alenljung}, \bibinfo{author}{J.~Lindblom},
  \bibinfo{author}{R.~Andreasson}, and \bibinfo{author}{T.~Ziemke},
  \enquote{\bibinfo{title}{User experience in social human-robot interaction}},
  in \emph{\bibinfo{booktitle}{Rapid automation: Concepts, methodologies,
  tools, and applications}}  (\bibinfo{publisher}{IGI Global},
  \bibinfo{year}{2019}), pp. \bibinfo{pages}{1468--1490}.

\bibitem{han2015two}
\bibinfo{author}{G.~Han}, \bibinfo{author}{L.~Wan}, \bibinfo{author}{L.~Shu},
  and \bibinfo{author}{N.~Feng}, \enquote{\bibinfo{title}{Two novel {DOA}
  estimation approaches for real-time assistant calibration systems in future
  vehicle industrial}},  \bibinfo{journal}{IEEE Syst. J.} \textbf{11}(3),
  \bibinfo{pages}{1361--1372} (\bibinfo{year}{2015}).

\bibitem{okutani2012outdoor}
\bibinfo{author}{K.~Okutani}, \bibinfo{author}{T.~Yoshida},
  \bibinfo{author}{K.~Nakamura}, and \bibinfo{author}{K.~Nakadai},
  \enquote{\bibinfo{title}{Outdoor auditory scene analysis using a moving
  microphone array embedded in a quadrocopter}}, in
  \emph{\bibinfo{booktitle}{IEEE Int. Conf. Intell. Robots Syst.}}
  (\bibinfo{year}{2012}), pp. \bibinfo{pages}{3288--3293}.

\bibitem{nakamura2011intelligent}
\bibinfo{author}{K.~Nakamura}, \bibinfo{author}{K.~Nakadai},
  \bibinfo{author}{F.~Asano}, and \bibinfo{author}{G.~Ince.},
  \enquote{\bibinfo{title}{Intelligent sound source localization and its
  application to multimodal human tracking}}, in \emph{\bibinfo{booktitle}{IEEE
  Intl. Conf. Intelligent Robots and Systems}} (\bibinfo{year}{2011}), pp.
  \bibinfo{pages}{143--148}.

\bibitem{wan2016application}
\bibinfo{author}{L.~Wan}, \bibinfo{author}{G.~Han}, \bibinfo{author}{L.~Shu},
  \bibinfo{author}{S.~Chan}, and \bibinfo{author}{T.~Zhu},
  \enquote{\bibinfo{title}{The application of {DOA} estimation approach in
  patient tracking systems with high patient density}},  \bibinfo{journal}{IEEE
  Trans. Ind. Electron.} \textbf{12}(6), \bibinfo{pages}{2353--2364}
  (\bibinfo{year}{2016}).

\bibitem{farmani2017informed}
\bibinfo{author}{M.~Farmani}, \bibinfo{author}{M.~S. Pedersen},
  \bibinfo{author}{Z.-H. Tan}, and \bibinfo{author}{J.~Jensen},
  \enquote{\bibinfo{title}{Informed sound source localization using relative
  transfer functions for hearing aid applications}},
  \bibinfo{journal}{IEEE/ACM Trans. Audio, Speech, Language Process.}
  \textbf{25}(3), \bibinfo{pages}{611--623} (\bibinfo{year}{2017}).

\bibitem{van1988beamforming}
\bibinfo{author}{B.~D.~V. Veen} and \bibinfo{author}{K.~M. Buckley},
  \enquote{\bibinfo{title}{Beamforming: A versatile approach to spatial
  filtering}},  \bibinfo{journal}{IEEE Acoustics, Speech, and Signal Processing
  Magazine} \textbf{5}(2), \bibinfo{pages}{4--24} (\bibinfo{year}{1988}).

\bibitem{music1986}
\bibinfo{author}{R.~Schmidt}, \enquote{\bibinfo{title}{Multiple emitter
  location and signal parameter estimation}},  \bibinfo{journal}{IEEE Trans.
  Antennas Propag.} \textbf{34}(3), \bibinfo{pages}{276--280}
  (\bibinfo{year}{1986}).

\bibitem{rao1989performance}
\bibinfo{author}{B.~D. Rao} and \bibinfo{author}{K.~Hari},
  \enquote{\bibinfo{title}{Performance analysis of root-music}},
  \bibinfo{journal}{IEEE Trans. on Acoustics, Speech, and Signal Proces.}
  \textbf{37}(12), \bibinfo{pages}{1939--1949} (\bibinfo{year}{1989}).

\bibitem{chen2001atomic}
\bibinfo{author}{S.~Chen}, \bibinfo{author}{D.~Donoho}, and
  \bibinfo{author}{M.~Saunders}, \enquote{\bibinfo{title}{Atomic decomposition
  by basis pursuit}},  \bibinfo{journal}{SIAM review} \textbf{43}(1),
  \bibinfo{pages}{129--159} (\bibinfo{year}{2001}).

\bibitem{compressivebeamforming}
\bibinfo{author}{A.~Xenaki}, \bibinfo{author}{P.~Gerstoft}, and
  \bibinfo{author}{K.~Mosegaard}, \enquote{\bibinfo{title}{Compressive
  beamforming}},  \bibinfo{journal}{J. Acoust. Soc. Am.} \textbf{136}(1),
  \bibinfo{pages}{260--271} (\bibinfo{year}{2014}).

\bibitem{l1svd}
\bibinfo{author}{D.~Malioutov}, \bibinfo{author}{M.~{\c{C}}etin}, and
  \bibinfo{author}{A.~S. Willsky}, \enquote{\bibinfo{title}{A sparse signal
  reconstruction perspective for source localization with sensor arrays}},
  \bibinfo{journal}{IEEE Trans. Signal Process} \textbf{53}(8),
  \bibinfo{pages}{3010--3022} (\bibinfo{year}{2005}).

\bibitem{omp}
\bibinfo{author}{J.~Tropp} and \bibinfo{author}{A.~C. Gilbert},
  \enquote{\bibinfo{title}{Signal recovery from partial information via
  orthogonal matching pursuit}},  \bibinfo{journal}{IEEE Trans. Inf. Theory}
  \textbf{53}(12), \bibinfo{pages}{4655--4666} (\bibinfo{year}{2007}).

\bibitem{pati1993orthogonal}
\bibinfo{author}{Y.~C. Pati}, \bibinfo{author}{R.~Rezaiifar}, and
  \bibinfo{author}{P.~Krishnaprasad}, \enquote{\bibinfo{title}{Orthogonal
  matching pursuit: Recursive function approximation with applications to
  wavelet decomposition}}, in \emph{\bibinfo{booktitle}{IEEE Asilomar conf. on
  sig., sys. and comput.}} (\bibinfo{year}{1993}), pp. \bibinfo{pages}{40--44}.

\bibitem{tipping2001}
\bibinfo{author}{M.~E. Tipping}, \enquote{\bibinfo{title}{Sparse {B}ayesian
  learning and the relevance vector machine}},  \bibinfo{journal}{J. Mach.
  Learn. Res.} \textbf{1}, \bibinfo{pages}{211--244} (\bibinfo{year}{2001}).

\bibitem{wipf2007}
\bibinfo{author}{D.~P. Wipf} and \bibinfo{author}{B.~D. Rao},
  \enquote{\bibinfo{title}{An empirical {B}ayesian strategy for solving the
  simultaneous sparse approximation problem}},  \bibinfo{journal}{IEEE Trans.
  Signal Process} \textbf{55}(7), \bibinfo{pages}{3704--3716}
  (\bibinfo{year}{2007}).

\bibitem{gerstoft2015multiple}
\bibinfo{author}{P.~Gerstoft}, \bibinfo{author}{A.~Xenaki}, and
  \bibinfo{author}{C.~F. Mecklenbr{\"a}uker}, \enquote{\bibinfo{title}{Multiple
  and single snapshot compressive beamforming}},  \bibinfo{journal}{J. Acoust.
  Soc. Am.} \textbf{138}(4), \bibinfo{pages}{2003--2014}
  (\bibinfo{year}{2015}).

\bibitem{multifreq2019sparse}
\bibinfo{author}{S.~Nannuru}, \bibinfo{author}{K.~L. Gemba},
  \bibinfo{author}{P.~Gerstoft}, \bibinfo{author}{W.~S. Hodgkiss}, and
  \bibinfo{author}{C.~F. Mecklenbr{\"a}uker}, \enquote{\bibinfo{title}{Sparse
  {B}ayesian learning with multiple dictionaries}},  \bibinfo{journal}{Signal
  Process.} \textbf{159}, \bibinfo{pages}{159--170} (\bibinfo{year}{2019}).

\bibitem{gemba2019robust}
\bibinfo{author}{K.~L. Gemba}, \bibinfo{author}{S.~Nannuru}, and
  \bibinfo{author}{P.~Gerstoft}, \enquote{\bibinfo{title}{Robust ocean acoustic
  localization with sparse {B}ayesian learning}},  \bibinfo{journal}{IEEE J.
  Sel. Topics Signal Process.} \textbf{13}(1), \bibinfo{pages}{49--60}
  (\bibinfo{year}{2019}).

\bibitem{MLEefficient}
\bibinfo{author}{Z.~Liu}, \bibinfo{author}{Z.~Huang}, and
  \bibinfo{author}{Y.~Zhou}, \enquote{\bibinfo{title}{An efficient maximum
  likelihood method for direction-of-arrival estimation via sparse {B}ayesian
  learning}},  \bibinfo{journal}{IEEE Trans. Wireless Commun.} \textbf{11}(10),
  \bibinfo{pages}{1--11} (\bibinfo{year}{2012}).

\bibitem{pandey2021sparse}
\bibinfo{author}{R.~Pandey}, \bibinfo{author}{S.~Nannuru}, and
  \bibinfo{author}{A.~Siripuram}, \enquote{\bibinfo{title}{Sparse {B}ayesian
  learning for acoustic source localization}}, in
  \emph{\bibinfo{booktitle}{IEEE Inter. Conf. Acous., Spe., Sig. Proces.}}
  (\bibinfo{year}{2021}), pp. \bibinfo{pages}{4670--4674}.

\bibitem{pandey2022experimental}
\bibinfo{author}{R.~Pandey}, \bibinfo{author}{S.~Nannuru}, and
  \bibinfo{author}{P.~Gerstoft}, \enquote{\bibinfo{title}{Experimental
  validation of wideband sbl models for doa estimation}}, in
  \emph{\bibinfo{booktitle}{IEEE Euro. Signal Proces. Conf.}}
  (\bibinfo{year}{2022}), pp. \bibinfo{pages}{219--223}.

\bibitem{krim1996two}
\bibinfo{author}{H.~Krim} and \bibinfo{author}{M.~Viberg},
  \enquote{\bibinfo{title}{Two decades of array signal processing research: the
  parametric approach}},  \bibinfo{journal}{IEEE Signal Processing Magazine}
  \textbf{13}(4), \bibinfo{pages}{67--94} (\bibinfo{year}{1996}).

\bibitem{vantrees2002}
\bibinfo{author}{H.~L.~V. Trees}, \emph{\bibinfo{title}{Optimum Array
  Processing (Detection, Estimation, and Modulation Theory, Part IV)}}
  (\bibinfo{publisher}{John Wiley \& Sons}, \bibinfo{year}{2002}).

\bibitem{mvdr2005}
\bibinfo{author}{R.~G. Lorenz} and \bibinfo{author}{S.~P. Boyd},
  \enquote{\bibinfo{title}{Robust minimum variance beamforming}},
  \bibinfo{journal}{IEEE Trans. Sig. Process.} \textbf{53}(5),
  \bibinfo{pages}{1684--1696} (\bibinfo{year}{2005}).

\bibitem{dibiase2000high}
\bibinfo{author}{J.~H. DiBiase}, \emph{\bibinfo{title}{A high-accuracy,
  low-latency technique for talker localization in reverberant environments
  using microphone arrays}}  (\bibinfo{publisher}{Brown University Providence,
  RI}, \bibinfo{year}{2000}).

\bibitem{diaz2020robust}
\bibinfo{author}{D.~Diaz-Guerra}, \bibinfo{author}{A.~Miguel}, and
  \bibinfo{author}{J.~R. Beltran}, \enquote{\bibinfo{title}{Robust sound source
  tracking using {SRP-PHAT} and 3{D} convolutional neural networks}},
  \bibinfo{journal}{IEEE/ACM Trans. Audio, Speech, Language Process.}
  \textbf{29}, \bibinfo{pages}{300--311} (\bibinfo{year}{2020}).

\bibitem{park2021sequential}
\bibinfo{author}{Y.~Park}, \bibinfo{author}{F.~Meyer}, and
  \bibinfo{author}{P.~Gerstoft}, \enquote{\bibinfo{title}{Sequential sparse
  {Bayesian} learning for time-varying direction of arrival}},
  \bibinfo{journal}{J. Acoust. Soc. Am.} \textbf{149}(3),
  \bibinfo{pages}{2089--2099} (\bibinfo{year}{2021}).

\bibitem{opochinsky2021deep}
\bibinfo{author}{R.~Opochinsky}, \bibinfo{author}{G.~Chechik}, and
  \bibinfo{author}{S.~Gannot}, \enquote{\bibinfo{title}{Deep ranking-based
  {DOA} tracking algorithm}}, in \emph{\bibinfo{booktitle}{Euro. Signal
  Process. Conf. (EUSIPCO)}} (\bibinfo{year}{2021}), pp.
  \bibinfo{pages}{1020--1024}.

\bibitem{pandey2022parametric}
\bibinfo{author}{R.~Pandey} and \bibinfo{author}{S.~Nannuru},
  \enquote{\bibinfo{title}{Parametric models for doa trajectory localization}},
  in \emph{\bibinfo{booktitle}{IEEE Inter. Conf. Acous., Spe., Sig. Proces.}}
  (\bibinfo{year}{2022}).

\bibitem{shreyas_TLDL}
\bibinfo{author}{S.~Jaiswal}, \bibinfo{author}{R.~Pandey}, and
  \bibinfo{author}{S.~Nannuru}, \enquote{\bibinfo{title}{Deep architecture for
  doa trajectory localization}}, in \emph{\bibinfo{booktitle}{IEEE Inter. Conf.
  Acous., Spe., Sig. Proces. (ICASSP)}} (\bibinfo{year}{2023}).

\bibitem{tang2013compressed}
\bibinfo{author}{G.~Tang}, \bibinfo{author}{B.~N. Bhaskar},
  \bibinfo{author}{P.~Shah}, and \bibinfo{author}{B.~Recht},
  \enquote{\bibinfo{title}{Compressed sensing off the grid}},
  \bibinfo{journal}{IEEE Trans. on Inform. The.} \textbf{59}(11),
  \bibinfo{pages}{7465--7490} (\bibinfo{year}{2013}).

\bibitem{xu2014precise}
\bibinfo{author}{W.~Xu}, \bibinfo{author}{J.~F. Cai}, \bibinfo{author}{V.~K.
  Mishra}, \bibinfo{author}{M.~Cho}, and \bibinfo{author}{A.~Kruger},
  \enquote{\bibinfo{title}{Precise semidefinite programming formulation of
  atomic norm minimization for recovering d-dimensional $(d\geq2)$ off-the-grid
  frequencies}}, in \emph{\bibinfo{booktitle}{IEEE Info. theo. and appli.
  workshop (ITA)}} (\bibinfo{year}{2014}), pp. \bibinfo{pages}{1--4}.

\bibitem{xenaki2015grid}
\bibinfo{author}{A.~Xenaki} and \bibinfo{author}{P.~Gerstoft},
  \enquote{\bibinfo{title}{Grid-free compressive beamforming}},
  \bibinfo{journal}{J. Acoust. Soc. Am.} \textbf{137}(4),
  \bibinfo{pages}{1923--1935} (\bibinfo{year}{2015}).

\bibitem{bhaskar2013atomic}
\bibinfo{author}{B.~N. Bhaskar}, \bibinfo{author}{G.~Tang}, and
  \bibinfo{author}{B.~Recht}, \enquote{\bibinfo{title}{Atomic norm denoising
  with applications to line spectral estimation}},  \bibinfo{journal}{IEEE
  Trans. Signal Process.} \textbf{61}(23), \bibinfo{pages}{5987--5999}
  (\bibinfo{year}{2013}).

\bibitem{chi2014compressive}
\bibinfo{author}{Y.~Chi} and \bibinfo{author}{Y.~Chen},
  \enquote{\bibinfo{title}{Compressive two-dimensional harmonic retrieval via
  atomic norm minimization}},  \bibinfo{journal}{IEEE Trans. on Sig. Proces.}
  \textbf{63}(4), \bibinfo{pages}{1030--1042} (\bibinfo{year}{2014}).

\bibitem{yang2017two}
\bibinfo{author}{Y.~Yang}, \bibinfo{author}{Z.~Chu}, \bibinfo{author}{Z.~Xu},
  and \bibinfo{author}{G.~Ping}, \enquote{\bibinfo{title}{Two-dimensional
  grid-free compressive beamforming}},  \bibinfo{journal}{J. Acoust. Soc. Am.}
  \textbf{142}(2), \bibinfo{pages}{618--629} (\bibinfo{year}{2017}).

\bibitem{yang2018resolution}
\bibinfo{author}{Y.~Yang}, \bibinfo{author}{Z.~Chu},
  \bibinfo{author}{G.~G.~Ping}, and \bibinfo{author}{Z.~Xu},
  \enquote{\bibinfo{title}{Resolution enhancement of two-dimensional grid-free
  compressive beamforming}},  \bibinfo{journal}{J. Acoust. Soc. Am.}
  \textbf{143}(6), \bibinfo{pages}{3860--3872} (\bibinfo{year}{2018}).

\bibitem{zhang2022efficient}
\bibinfo{author}{Y.~Zhang}, \bibinfo{author}{Y.~Wang},
  \bibinfo{author}{Z.~Tian}, \bibinfo{author}{G.~Leus}, and
  \bibinfo{author}{G.~Zhang}, \enquote{\bibinfo{title}{Efficient
  super-resolution two-dimensional harmonic retrieval with multiple measurement
  vectors}},  \bibinfo{journal}{IEEE Trans. on Sig. Proces.} \textbf{70},
  \bibinfo{pages}{1224--1240} (\bibinfo{year}{2022}).

\bibitem{yang2016vandermonde}
\bibinfo{author}{Z.~Yang}, \bibinfo{author}{L.~Xie}, and
  \bibinfo{author}{P.~Stoica}, \enquote{\bibinfo{title}{Vandermonde
  decomposition of multilevel toeplitz matrices with application to
  multidimensional super-resolution}},  \bibinfo{journal}{IEEE Trans. on Inf.
  Theory} \textbf{62}(6), \bibinfo{pages}{3685--3701} (\bibinfo{year}{2016}).

\bibitem{wu2022maximum}
\bibinfo{author}{X.~Wu}, \bibinfo{author}{Z.~Yang},
  \bibinfo{author}{P.~Stoica}, and \bibinfo{author}{Z.~Xu},
  \enquote{\bibinfo{title}{Maximum likelihood line spectral estimation in the
  signal domain: A rank-constrained structured matrix recovery approach}},
  \bibinfo{journal}{IEEE Trans. on Sig. Process.} \textbf{70},
  \bibinfo{pages}{4156--4169} (\bibinfo{year}{2022}).

\bibitem{semper2018grid}
\bibinfo{author}{S.~Semper}, \bibinfo{author}{F.~Roemer},
  \bibinfo{author}{T.~Hotz}, and \bibinfo{author}{G.~D. Galdo},
  \enquote{\bibinfo{title}{Grid-free direction-of-arrival estimation with
  compressed sensing and arbitrary antenna arrays}}, in
  \emph{\bibinfo{booktitle}{IEEE Inter. Conf. Acous., Spe., Sig. Proces.
  (ICASSP)}} (\bibinfo{year}{2018}), pp. \bibinfo{pages}{3251--3255}.

\bibitem{wagner2021gridless}
\bibinfo{author}{M.~Wagner}, \bibinfo{author}{Y.~Park}, and
  \bibinfo{author}{P.~Gerstoft}, \enquote{\bibinfo{title}{Gridless doa
  estimation and root-music for non-uniform linear arrays}},
  \bibinfo{journal}{IEEE Trans. on Sig. Proces.} \textbf{69},
  \bibinfo{pages}{2144--2157} (\bibinfo{year}{2021}).

\bibitem{wu2022gridless}
\bibinfo{author}{Y.~Wu}, \bibinfo{author}{M.~B. Wakin}, and
  \bibinfo{author}{P.~Gerstoft}, \enquote{\bibinfo{title}{Gridless doa
  estimation with multiple frequencies}},  \bibinfo{journal}{arXiv preprint
  arXiv:2207.06159}  (\bibinfo{year}{2022}).

\bibitem{jiang2020gridless}
\bibinfo{author}{Y.~Jiang}, \bibinfo{author}{D.~Li}, \bibinfo{author}{X.~Wu},
  and \bibinfo{author}{W.~P. Zhu}, \enquote{\bibinfo{title}{A gridless wideband
  doa estimation based on atomic norm minimization}}, in
  \emph{\bibinfo{booktitle}{Sen. Arr. and Multi. Sig. Proces. Work. (SAM)}},
  \bibinfo{organization}{IEEE} (\bibinfo{year}{2020}), pp.
  \bibinfo{pages}{1--5}.

\bibitem{ang2020multiband}
\bibinfo{author}{Y.~Y. Ang}, \bibinfo{author}{N.~Nguyen}, and
  \bibinfo{author}{W.~S. Gan}, \enquote{\bibinfo{title}{Multiband grid-free
  compressive beamforming}},  \bibinfo{journal}{Mech. Sys. and Sig. Proces.}
  \textbf{135}, \bibinfo{pages}{106425} (\bibinfo{year}{2020}).

\bibitem{chardon2021gridless}
\bibinfo{author}{G.~Chardon} and \bibinfo{author}{U.~Boureau},
  \enquote{\bibinfo{title}{Gridless three-dimensional compressive beamforming
  with the sliding frank-wolfe algorithm}},  \bibinfo{journal}{J. Acoust. Soc.
  Am.} \textbf{150}(4), \bibinfo{pages}{3139--3148} (\bibinfo{year}{2021}).

\bibitem{mamandipoor2016newtonized}
\bibinfo{author}{B.~Mamandipoor}, \bibinfo{author}{D.~Ramasamy}, and
  \bibinfo{author}{U.~Madhow}, \enquote{\bibinfo{title}{Newtonized orthogonal
  matching pursuit: Frequency estimation over the continuum}},
  \bibinfo{journal}{IEEE Trans. on Sig. Proces.} \textbf{64}(19),
  \bibinfo{pages}{5066--5081} (\bibinfo{year}{2016}).

\bibitem{denoyelle2019sliding}
\bibinfo{author}{Q.~Denoyelle}, \bibinfo{author}{V.~Duval},
  \bibinfo{author}{G.~Peyr{\'e}}, and \bibinfo{author}{E.~Soubies},
  \enquote{\bibinfo{title}{The sliding frank--wolfe algorithm and its
  application to super-resolution microscopy}},  \bibinfo{journal}{Inverse
  Problems} \textbf{36}(1), \bibinfo{pages}{014001} (\bibinfo{year}{2019}).

\bibitem{Beurling2012exact}
\bibinfo{author}{Y.~D. Castro} and \bibinfo{author}{F.~Gamboa},
  \enquote{\bibinfo{title}{Exact reconstruction using beurling minimal
  extrapolation}},  \bibinfo{journal}{Journal of Mathematical Analysis and
  applications} \textbf{395}(1), \bibinfo{pages}{336--354}
  (\bibinfo{year}{2012}).

\bibitem{gerstoft2016}
\bibinfo{author}{P.~Gerstoft}, \bibinfo{author}{C.~F. Mecklenbr{\"a}uker},
  \bibinfo{author}{A.~Xenaki}, and \bibinfo{author}{S.~Nannuru},
  \enquote{\bibinfo{title}{Multisnapshot sparse {B}ayesian learning for
  {DOA}}},  \bibinfo{journal}{IEEE Signal Process. Lett.} \textbf{23}(10),
  \bibinfo{pages}{1469--1473} (\bibinfo{year}{2016}).

\bibitem{tabaghi2019kinetic}
\bibinfo{author}{P.~Tabaghi}, \bibinfo{author}{I.~Dokmani{\'c}}, and
  \bibinfo{author}{M.~Vetterli}, \enquote{\bibinfo{title}{Kinetic euclidean
  distance matrices}},  \bibinfo{journal}{IEEE Trans. on Sig. Proces.}
  \textbf{68}, \bibinfo{pages}{452--465} (\bibinfo{year}{2019}).

\bibitem{mallat1993matching}
\bibinfo{author}{S.~G. Mallat} and \bibinfo{author}{Z.~Zhang},
  \enquote{\bibinfo{title}{Matching pursuits with time-frequency
  dictionaries}},  \bibinfo{journal}{IEEE Trans. on sig. proces.}
  \textbf{41}(12), \bibinfo{pages}{3397--3415} (\bibinfo{year}{1993}).

\bibitem{cai2011orthogonal}
\bibinfo{author}{T.~Cai} and \bibinfo{author}{L.~Wang},
  \enquote{\bibinfo{title}{Orthogonal matching pursuit for sparse signal
  recovery with noise}},  \bibinfo{journal}{IEEE Trans. on Info. theo.}
  \textbf{57}(7), \bibinfo{pages}{4680--4688} (\bibinfo{year}{2011}).

\bibitem{nocedal1999numerical}
\bibinfo{author}{J.~Nocedal} and \bibinfo{author}{S.~J. Wright},
  \emph{\bibinfo{title}{Numerical optimization}}
  (\bibinfo{publisher}{Springer}, \bibinfo{year}{1999}).

\bibitem{yang2020two}
\bibinfo{author}{Y.~Yang}, \bibinfo{author}{Z.~Chu}, \bibinfo{author}{Y.~Yang},
  and \bibinfo{author}{S.~Yin}, \enquote{\bibinfo{title}{Two-dimensional
  newtonized orthogonal matching pursuit compressive beamforming}},
  \bibinfo{journal}{J. Acoust. Soc. Am.} \textbf{148}(3),
  \bibinfo{pages}{1337--1348} (\bibinfo{year}{2020}).

\bibitem{bohme1985source}
\bibinfo{author}{J.~Bohme}, \enquote{\bibinfo{title}{Source-parameter
  estimation by approximate maximum likelihood and nonlinear regression}},
  \bibinfo{journal}{IEEE J. Ocean. Eng.} \textbf{10}(3),
  \bibinfo{pages}{206--212} (\bibinfo{year}{1985}).

\bibitem{schuhmacher2008consistent}
\bibinfo{author}{D.~Schuhmacher}, \bibinfo{author}{B.-T. Vo}, and
  \bibinfo{author}{B.-N. Vo}, \enquote{\bibinfo{title}{A consistent metric for
  performance evaluation of multi-object filters}},  \bibinfo{journal}{IEEE
  Trans. on Sig. Proces.} \textbf{56}(8), \bibinfo{pages}{3447--3457}
  (\bibinfo{year}{2008}).

\bibitem{ospa_2}
\bibinfo{author}{B.~Ristic}, \bibinfo{author}{B.-N. Vo},
  \bibinfo{author}{D.~Clark}, and \bibinfo{author}{B.~T. Vo},
  \enquote{\bibinfo{title}{A metric for performance evaluation of multi-target
  tracking algorithms}},  \bibinfo{journal}{IEEE Trans. on Sig. Proces.}
  \textbf{59}(7), \bibinfo{pages}{3452--3457} (\bibinfo{year}{2011}).

\end{thebibliography}
\end{document}